# Fast Maximum-Likelihood Decoder for 4×4 Quasi-Orthogonal Space-Time Block Code

Adel Ahmadi, *Student Member, IEEE*, Siamak Talebi, *Member, IEEE*

*Abstract*—This letter introduces two fast maximum-likelihood (ML) detection methods for 4×4 quasi-orthogonal space-time block code (QOSTBC). The first algorithm with a relatively simple design exploits structure of quadrature amplitude modulation (QAM) constellations to achieve its goal and the second algorithm, though somewhat more complex, can be applied to any arbitrary constellation. Both decoders utilize a novel decomposition technique for ML metric which divides the metric into independent positive parts and a positive interference part. Search spaces of symbols are substantially reduced by employing the independent parts and statistics of noise. Finally, the members of search spaces are successively evaluated until the metric is minimized. Simulation results confirm that the proposed decoder is superior to some of the most recently published methods in terms of complexity level. More specifically, the results verified that application of the new algorithm with 1024-QAM would require reduced computational complexity compared to state-of-the-art solution with 16-QAM.

## I. INTRODUCTION

QUASI-ORTHOGONAL space-time block codes (QOSTBCs) are noteworthy nowadays due to their desired performances and pairwise detections. However, complexity of pairwise maximum-likelihood (ML) decoder rises drastically when size of constellation is increased. A survey of related literature shows that much research has been underway to develop an approach that reduces complexity of QOSTBC decoder [1]-[7]. Authors of [1] employ QR decomposition and sorting to simplify detection of QOSTBC with four transmit antennas. This method achieves ML performance and offers low complexity. QOSTBCs with minimum decoding complexity are proposed in [2] with degraded error performance compared to conventional QOSTBC. A suboptimum decoder is presented in [3] based on sorted QR decomposition and real-valued representation of [2]. The decoder unveiled in [3] offers near-ML error performance and its complexity is independent of constellation size. In [4], a low complexity ML decoder is introduced on the basis of QR and real valued lattice representation. For rotated QOSTBC with four transmit antennas and quadrature amplitude modulation (QAM) constellation, [5] explores a fast scheme with ML error performance and reduced complexity. In [6], a different structure is proposed for QOSTBCs decoding with three or four transmits antennas. This method achieves a near-ML performance with low complexity decoder. A new 4×4 QOSTBC is reported in [7] that employs precoder and rotated symbols. The complexity of decoder is low and a reasonable suboptimum error performance is obtained when rotation angles are optimized. In [8], a suboptimum fast decoding algorithm is investigated for block diagonal QOSTBC with arbitrary transmit antennas.

The innovative methods developed in this letter decompose the received vector into two pairs of symbols which are detected independently. To do this, the ML metric minimization of each pair is transformed into sum of independent positive parts and an interference part. It should be noted that the independency between parts facilitates detection by helping to considerably limit search spaces of the symbols. The candidates placed in relevant partial search area are gradually evaluated and then transmitted symbols are estimated by computing interference between them. If the search areas are small and no symbol is detected, then they are extended and evaluation is repeated till transmitted symbols are detected. The novel ML metric decomposition studied in this letter boasts two important features, namely; decomposition is not a highly complex process, and more significantly, most of the decomposed parts are independent of each other. The first proposed method utilizes the structure of QAM to further reduce complexity of detection and the second method, which is relatively more complex, can be applied to any arbitrary constellations. Both of these new algorithms offer the desired ML performance.

The rest of the letter is organized as follows. Section II describes a wireless communication system based on 4×4 QOSTBC. The proposed fast ML detection methods are divulged in Section III. This is followed by Section IV which briefly covers simulation set up and compares the results against those of other fast methods. The letter ends with a conclusion in Section V.

## II. QUASI-ORTHOGONAL SPACE TIME BLOCK CODE

QOSTBCs have the following form for $M_T = 4$ transmit

A. Ahmadi is with the Department of Electrical Engineering, Shahid Bahonar University of Kerman, Kerman, Iran (e-mail: adel.ahmadi@ieee.org).

S. Talebi is with the Department of Electrical Engineering, Shahid Bahonar University of Kerman, Kerman, Iran and the Advanced Communications Research Institute, Sharif University, Tehran, Iran (e-mail: siamak.talebi@uk.ac.ir).

antennas:
$$X = \begin{bmatrix} s_1 & s_2 & \tilde{s}_3 & \tilde{s}_4 \\ -s_2^* & s_1^* & -\tilde{s}_4^* & \tilde{s}_3^* \\ \tilde{s}_3 & \tilde{s}_4 & s_1 & s_2 \\ -\tilde{s}_4^* & \tilde{s}_3^* & -s_2^* & s_1^* \end{bmatrix}, \quad (1)$$

where $\tilde{s}_3 = e^{j\theta}s_3$, $\tilde{s}_4 = e^{j\theta}s_4$ and the data symbols $s_1, \ldots, s_4$ belong to constellation $\mathcal{C}$. A space-time communication system that transmits four symbols over four time slots can be equivalently represented for $j$th receive antenna as:

$$y_j = \sqrt{\frac{\rho}{M_T}} H_j s + w_j, \quad (2)$$

where $j = 1,2,\ldots,M_R$, $y_j = [y_{1,j}, \ldots, y_{M_T,j}]^T$ and $w_j = [w_{1,j}, \ldots, w_{M_T,j}]^T$. The equivalent received vector, equivalent additive white Gaussian noise (AWGN) vector and equivalent channel matrices are represented by $w_j$, $y_j$ and $H_j$, respectively. The signal-to-noise ratio (SNR) is indicated by $\rho$, and $y_{i,j}$ and $w_{i,j}$ denote equivalent received signal and noise at the $i$th time slot of the $j$th receive antenna, respectively. The AWGN has complex Gaussian distribution with zero mean and unit variance. The equivalent channel for the $j$th receive antenna can be defined as:

$$H_j = \begin{bmatrix} \mathcal{B}(h_{1,j}, h_{2,j}) & e^{j\theta}\mathcal{B}(h_{3,j}, h_{4,j}) \\ \mathcal{B}(h_{3,j}, h_{4,j}) & e^{j\theta}\mathcal{B}(h_{1,j}, h_{2,j}) \end{bmatrix}, \quad (3)$$

where $h_{i,j}$ stands for channel fade between the $i$th transmit antenna and the $j$th receive antenna such that:

$$\mathcal{B}(h_{i,j}, h_{i+1,j}) \triangleq \begin{bmatrix} h_{i,j} & h_{i+1,j} \\ h_{i+1,j}^* & -h_{i,j}^* \end{bmatrix}. \quad (4)$$

The complete received vectors can be concatenated as:

$$y = \sqrt{\frac{\rho}{M_T}} H s + w, \quad (5)$$

where $\triangleq [y_1^T, \ldots, y_{M_R}^T]^T$, $w \triangleq [w_1^T, \ldots, w_{M_R}^T]^T$ and $H \triangleq [H_1^T, \ldots, H_{M_R}^T]^T$.

## III. Fast Maximum-Likelihood Decoder

The ML decoder should minimize the following norm in order to estimate the transmitted symbols:

$$\tilde{s} = \arg\min_{s \in \mathcal{C}^{M_T}} \left\| y - \sqrt{\frac{\rho}{M_T}} H s \right\|. \quad (6)$$

The above minimization can be rewritten as:

$$\tilde{s} = \arg\min_{s \in \mathcal{C}^{M_T}} \|H(s - z)\|, \quad (7)$$

where $z = [z_1, \ldots, z_{M_T}]$ is defined as:

$$z \triangleq \sqrt{\frac{M_T}{\rho}} (H^H H)^{-1} H^H y. \quad (8)$$

The matrix $H_j$ can be decomposed into:

$$H_j = \frac{1}{2} U H_j' U V, \quad (9)$$

where $V \triangleq \mathrm{diag}(I_2, e^{j\theta}I_2,)$,

$$U \triangleq \begin{bmatrix} I_2 & I_2 \\ I_2 & -I_2 \end{bmatrix}, \quad (10)$$

$$H_j' \triangleq \begin{bmatrix} \mathcal{B}(h_{13,j}^+, h_{24,j}^+) & 0_{2\times 2} \\ 0_{2\times 2} & \mathcal{B}(h_{13,j}^-, h_{24,j}^-) \end{bmatrix}. \quad (11)$$

Further, the 2×2 identity matrix is represented by $I_2$ and $h_{13,j}^+ \triangleq h_{1,j} + h_{3,j}$, $h_{13,j}^- \triangleq h_{1,j} - h_{3,j}$, $h_{24,j}^+ \triangleq h_{2,j} + h_{4,j}$, and $h_{24,j}^- \triangleq h_{2,j} - h_{4,j}$.

By employing (9)-(11) and doing some math operations, we are able to simplify $T \triangleq H^H H$ as:

$$T = \begin{bmatrix} t_0 I_2 & e^{j\theta}t_1 I_2 \\ e^{-j\theta}t_1 I_2 & t_0 I_2 \end{bmatrix}, \quad (12)$$

where $t_0 = (\alpha + \beta)/2$, $t_1 = (\alpha - \beta)/2$ and

$$\alpha = \sum_{j=1}^{N} \left( |h_{13,j}^+|^2 + |h_{24,j}^+|^2 \right), \quad (13)$$

$$\beta = \sum_{j=1}^{N} \left( |h_{13,j}^-|^2 + |h_{24,j}^-|^2 \right). \quad (14)$$

On the other hand, $z$ can be computed with fewer multiplications by utilizing the above relations, yielding

$$z = \sqrt{\frac{M_T}{4\rho}} \begin{bmatrix} \frac{1}{\alpha} I_2 & \frac{1}{\beta} I_2 \\ \frac{e^{-j\theta}}{\alpha} I_2 & -\frac{e^{-j\theta}}{\beta} I_2 \end{bmatrix} \sum_{j=1}^{N} H_j'^H (U y_j). \quad (15)$$

By substituting (12) into $\|H(s - z)\|^2$, we obtain

$$(s_{13} - z_{13})^H T'(s_{13} - z_{13}) + (s_{24} - z_{24})^H T'(s_{24} - z_{24}), \quad (16)$$

where $z_{13} \triangleq [z_1, z_3]^T$, $z_{24} \triangleq [z_2, z_4]^T$, $s_{13} \triangleq [s_1, s_3]^T$, $s_{24} \triangleq [s_2, s_4]^T$, and

$$T' = \begin{bmatrix} t_0 & e^{j\theta}t_1 \\ e^{-j\theta}t_1 & t_0 \end{bmatrix}. \quad (17)$$

Based on (16), the ML decoder can detect two pairs $(s_1, s_3)$ and $(s_2, s_4)$ independently:

$$(\tilde{s}_1, \tilde{s}_3) = \arg\min_{s_1, s_3 \in \mathcal{C}} \{(s_{13} - z_{13})^H T'(s_{13} - z_{13})\}, \quad (18)$$
$$(\tilde{s}_2, \tilde{s}_4) = \arg\min_{s_2, s_4 \in \mathcal{C}} \{(s_{24} - z_{24})^H T'(s_{24} - z_{24})\}. \quad (19)$$

For the remaining part of this section, we focus on detection of $(s_1, s_3)$ noting that the other pair $(s_2, s_4)$ can be detectable by applying a similar approach.

The ML metric of (18) can be expanded as:

$$t_0(|s_1 - z_1|^2 + |s_3 - z_3|^2) + 2t_1 \Re\{e^{j\theta}(s_1 - z_1)^*(s_3 - z_3)\}, \quad (20)$$

where $\Re(z)$ denotes the real part of $z$. Based on (13) and (14), $t_0 \geq |t_1|$ and therefore we can rewrite minimization of (20) as minimization of sum of three positive parts:

$$|s_1 - z_1|^2 + |s_3 - z_3|^2 + t_2 \left| e^{-j\theta/2}(s_1 - z_1) + \mathrm{sign}(t_1)e^{j\theta/2}(s_3 - z_3) \right|^2, \quad (21)$$

where $t_2 \triangleq |t_1|/(t_0 - |t_1|)$ and $\mathrm{sign}(\cdot)$ stands for the signum function. In (21), the first part $|s_1 - z_1|^2$ and the second part $|s_3 - z_3|^2$ are independent of each other which helps to reduce the search space, and the purpose of the third part is to present interference between the other parts and thus lead to detection of $\tilde{s}_1$ and $\tilde{s}_3$.

In subsection A, a fast ML decoder is presented for QAM constellations and in subsection B a fast ML method is introduced that deals with arbitrary constellations. For the proposed algorithms, algorithm initialization and detection complexity are covered in subsections C and D, respectively.

### A. Fast ML Detection for QAM Constellations

In this subsection, we investigate detection of QAM constellations. For the sake of simplicity and clarity, let consider a square $M^2$-QAM constellation but the solution can also be straightforwardly extended to any rectangular QAM constellations. Under this scenario, the real and imaginary parts of symbols belong to $\mathcal{R} = \{p_m \in \mathbb{R}, m = 1,2,\ldots,M\}$.

Two independent parts of (21) can be expanded to a summation of their real and imaginary parts:

$$\sum_{k=1,3}\{(\dot{s}_k - \dot{z}_k)^2 + (\ddot{s}_k - \ddot{z}_k)^2\}, \quad (22)$$

where $\dot{z}_k$ and $\ddot{z}_k$ stand for the real and imaginary parts of $z_k$, respectively. The decoder searches within $\mathcal{R}$ to find the best choices (i.e. $\tilde{\dot{s}}_k$ and $\tilde{\ddot{s}}_k$) for $\dot{s}_k$ and $\ddot{s}_k$, when $k = 1,3$. By analyzing (22) and assuming that the minimum of (21) is smaller than $r^2$, we have:

$$|\dot{s}_k - \dot{z}_k| < r, \quad (23)$$
$$|\ddot{s}_k - \ddot{z}_k| < r, \quad (24)$$

where $k = 1,3$. The above inequalities help us to discard inappropriate members of $\mathcal{R}$ by comparing them with the real and imaginary parts of $z_1$ and $z_3$. Therefore, the decoder initially selects certain members of $\mathcal{R}$, which are located within intervals $[\dot{z}_k - r, \dot{z}_k + r]$ and $[\ddot{z}_k - r, \ddot{z}_k + r]$, when $k = 1,3$. Then, the selected positions are evaluated step-by-step to extract $\tilde{s}_1 = \tilde{\dot{s}}_1 + j\tilde{\ddot{s}}_1$ and $\tilde{s}_3 = \tilde{\dot{s}}_3 + j\tilde{\ddot{s}}_3$ which minimize the ML metric.

The proposed algorithm that detects $(\tilde{s}_1, \tilde{s}_3)$ for an $M^2$-QAM can be summarized as follows:

**Step 1**: Compute $t_0$, $t_1$ and $\mathbf{z}$ by employing (10)-(15) and adjust $r$ to an appropriate value, then define $\lambda \triangleq t_1/t_0$, $\tau_1 \triangleq 2\lambda \cos\theta$, $\tau_2 \triangleq 2\lambda \sin\theta$ and $\vartheta \triangleq (1 - |\lambda|)^{-1}$.

**Step 2**: Define $\mathbf{z}' \in \mathbb{R}^{4\times 1}$ as $\mathbf{z}' \triangleq [\dot{z}_1, \ddot{z}_1, \dot{z}_3, \ddot{z}_3]^T$.

**Step 3**: Adjust $\gamma^2$ to $\gamma^2 = (1 - |\lambda|)r^2$.

**Step 4**: Select a new value of $p_1$ from $\mathcal{R}$, while it is located inside of interval $[z'_1 - r, z'_1 + r]$; If there isn't any new point go to Step 12.

**Step 5**: Set $n = 2$, $\delta_1 = p_1 - z'_1$ and $\varepsilon_1 = \delta_1^2$.

**Step 6**: Select a new value for $p_n$ from $\mathcal{R}$ and inside of interval $[z'_n - r, z'_n + r]$; If there isn't any new point, modify $n$ as $n = n - 2$ and go to Step 8.

**Step 7**: Set $\delta_n = p_n - z'_n$ and $\varepsilon_n = \varepsilon_{n-1} + \delta_n^2$. If $\varepsilon_n > r^2$, go back to Step 6.

**Step 8**: If $1 \leq n \leq 3$, update $n$ as $n = n + 1$ and return to Step 6. If $n = 0$, return to Step 4.

**Step 9**: Compute value of ML metric by utilizing $\varepsilon = \varepsilon_4 + \tau_1\kappa_1 - \tau_2\kappa_2$, where $\kappa_1 = \delta_1\delta_3 + \delta_2\delta_4$ and $\kappa_2 = \delta_1\delta_4 - \delta_2\delta_3$.

**Step 10**: If $\varepsilon < \gamma^2$, set $\gamma^2 = \varepsilon$, $r^2 = \vartheta\varepsilon$, $\tilde{s}_1 = p_1 + jp_2$ and $\tilde{s}_3 = p_3 + jp_4$.

**Step 11**: Go back to Step 6.

**Step 12**: If $\tilde{s}_1$ and $\tilde{s}_3$ are not still obtained, increase the values of $r^2$ and $\gamma^2$ and repeat Steps 3 to 12; Otherwise, the final result of detection is $(\tilde{s}_1, \tilde{s}_3)$.

In the following subsection, a fast ML detector is examined for arbitrary constellations.

### B. Fast ML Detection of Arbitrary Constellations

For arbitrary constellation $\mathcal{C}$ with $M$ points, we cannot exploit lattice structure and therefore decoding calls for primary form of (20) and (21). The proposed technique can be summarized as follows:

**Step 1**: Compute $t_0$, $t_1$ and $\mathbf{z}$ by employing (10)-(15) and adjust $r$ to an appropriate value, then define $\lambda \triangleq t_1/t_0$, $\tau_1 \triangleq 2\lambda \cos\theta$, $\tau_2 \triangleq 2\lambda \sin\theta$ and $\vartheta \triangleq (1 - |\lambda|)^{-1}$.

**Step 2**: Adjust $\gamma^2$ to $\gamma^2 = (1 - |\lambda|)r^2$.

**Step 3**: Select a new complex value of $\dot{c}_1 + j\ddot{c}_1$ from $\mathcal{C}$, whose real and imaginary parts are located within $[\dot{z}_k - r, \dot{z}_k + r]$ and $[\ddot{z}_k - r, \ddot{z}_k + r]$, respectively. If there isn't any new point, go to Step 11.

**Step 4**: Set $\dot{\delta}_1 = \dot{c}_1 - \dot{z}_1$, $\ddot{\delta}_2 = \ddot{c}_1 - \ddot{z}_1$ and $\varepsilon_1 = \dot{\delta}_1^2 + \ddot{\delta}_1^2$. If $\varepsilon_1 > r^2$, go to the previous step and select another point.

**Step 5**: From $\mathcal{C}$, select a new complex value of $\dot{c}_3 + j\ddot{c}_3$, whose real and imaginary parts are located within $[\dot{z}_3 - r, \dot{z}_3 + r]$ and $[\ddot{z}_3 - r, \ddot{z}_3 + r]$, respectively. If there isn't any new point, go to Step 3.

**Step 6**: Set $\dot{\delta}_3 = \dot{c}_3 - \dot{z}_3$ and $\varepsilon_3 = \varepsilon_1 + \dot{\delta}_3^2$. If $\varepsilon_3 > r^2$, go to the previous step and select another point.

**Step 7**: Set $\ddot{\delta}_3 = \ddot{c}_3 - \ddot{z}_3$ and $\varepsilon_3 = \varepsilon_3 + \ddot{\delta}_3^2$. If $\varepsilon_3 > r^2$, go to Step 5 and select another point.

**Step 8**: Compute value of ML metric by utilizing $\varepsilon = \varepsilon_3 + \tau_1\kappa_1 - \tau_2\kappa_2$, where $\kappa_1 = \dot{\delta}_1\dot{\delta}_3 + \ddot{\delta}_2\ddot{\delta}_3$ and $\kappa_2 = \dot{\delta}_1\ddot{\delta}_3 - \ddot{\delta}_2\dot{\delta}_3$.

**Step 9**: If $\varepsilon < \gamma^2$, set $\gamma^2 = \varepsilon$, $r^2 = \vartheta\varepsilon$, $\tilde{s}_1 = \dot{c}_1 + j\ddot{c}_1$ and $\tilde{s}_3 = \dot{c}_3 + j\ddot{c}_3$.

**Step 10**: Go back to Step 5.

**Step 11**: If $\tilde{s}_1$ and $\tilde{s}_3$ are not still obtained, increase the values of $r^2$ and $\gamma^2$ and again repeat Steps 3 to 11; Otherwise, the final result of detection is $(\tilde{s}_1, \tilde{s}_3)$.

Next, selection of $r$ and complexity of the proposed method are studied.

### C. Algorithm Initialization

The proposed methods need to begin with an appropriate initial value for $r$ which should be suitably selected to avoid unnecessary complexity. From the definition (8), $\mathbf{z}$ can be rewritten as:

$$\mathbf{z} = \mathbf{s} + \boldsymbol{\omega}, \quad (25)$$

where $\mathbf{s} = [s_1, \ldots, s_4]^T$ is the transmitted vector and $\boldsymbol{\omega} = [\omega_1, \ldots, \omega_4]^T$ is complex Gaussian noise vector:

$$\boldsymbol{\omega} = \sqrt{\frac{M_T}{\rho}}\mathbf{T}^{-1}\mathbf{H}^H\mathbf{w}. \quad (26)$$

The mean of $\boldsymbol{\omega}$ is $\mathbf{0}_{4\times 1}$ and its covariance can be simplified as:

$$\mathbf{R}_\omega = \begin{bmatrix} \sigma_1^2\mathbf{I}_2 & e^{j\theta}\varrho\mathbf{I}_2 \\ e^{-j\theta}\varrho\mathbf{I}_2 & \sigma_1^2\mathbf{I}_2 \end{bmatrix}, \quad (27)$$

where $\sigma_1^2 = M_T(\alpha + \beta)/2\rho\alpha\beta$ and $\varrho = M_T(\alpha - \beta)/2\rho\alpha\beta$. If the ML decoder correctly identifies the transmitted symbols, then minimum of ML has occurred for transmitted symbols. Under this condition, the part $\dot{s}_i - \dot{z}_i$ of minimum metric becomes equal to $-\dot{\omega}_i$ for $i = 1, \ldots, 4$. Most of the times, the absolute of $\dot{\omega}_i$ is smaller than its standard deviation $\sigma_1/2$ multiplied by four and therefore $r = 2\sigma_1$ is an appropriate choice for initializing.

On the other hand, $r$ can also be expressed as $r = d_{min}/2$ where $d_{min}$ is the minimum distance between two distinct constellation points. In this letter, $r = d_{min}/2$ is used for simulation because it involves lower level of complexity.

### D. Complexity of proposed methods

The decoder follows a procedure that can be divided into pre-computation and search stages. The initial value of variables and candidate points are obtained by engaging about $8M + 1$ relational operator, $64M_R + 29$ real additions, $40M_R + 31$ multiplications and 4 divisions during the first stage. The complexity of the procedure in the second stage depends on AWGN which practically becomes smaller than $d_{min}$ when SNR is sufficiently high. Under this condition, on average, one or two candidates are selected for evaluation per each dimension requiring about 22-44 additions and 22-44 multiplications for a QAM constellation. Hence, complexity of the search stage is low and that of pre-computation is taken as the dominant factor.

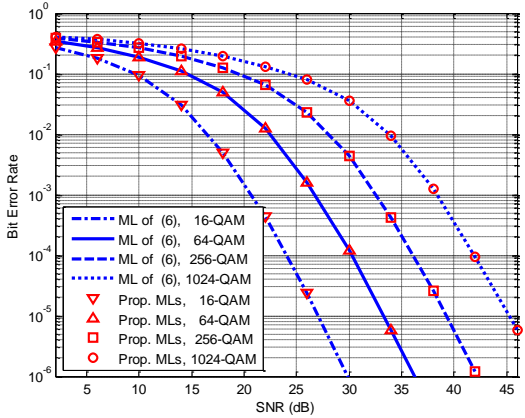

Fig. 1. Error performance of proposed method: BER of proposed method is same as that of ML/sphere.

TABLE I
AVERAGE NUMBER OF ADDITIONS AND MULTIPLICATIONS FOR BER$\approx 10^{-4}$

|  | 16-QAM | | 64-QAM | | 256-QAM | | 1024-QAM | |
| --- | --- | --- | --- | --- | --- | --- | --- | --- |
| Method | + | × | + | × | + | × | + | × |
| Prop. I | 116.2 | 94.1 | 116.8 | 94.7 | 117.6 | 95.4 | 118.0 | 95.7 |
| Prop. II | 116.8 | 94.4 | 118.1 | 95.4 | 119.4 | 96.4 | 119.8 | 96.7 |

TABLE II
AVERAGE NUMBER OF ADDITIONS AND MULTIPLICATIONS

|  | 16-QAM | | 64-QAM | | 256-QAM | |
| --- | --- | --- | --- | --- | --- | --- |
| Method | + | × | + | × | + | × |
| [1] | 373 | 469 | 1311 | 1179 | 5637 | 4302 |
| [2] | 1728 | 2048 | 6912 | 8192 | 27648 | 32768 |
| [3] | 388 | 544 | 388 | 544 | 388 | 544 |
| [4] | 656 | 592 | 2192 | 1744 | 8336 | 6352 |
| [5] | 327 | 437 | 647 | 443 | 1479 | 1445 |
| [6] | 232 | 308 | 808 | 1076 | 3112 | 4148 |
| [7] | 225 | 332 | 225 | 348 | 225 | 364 |

Compared against the sphere method, where the pre-computation stage requires about $424M_R + 28$ additions, $432M_R + 120$ multiplications with real equivalent channel having $8M_R$ rows and 8 columns [9], or against the QR-based methods involving about $104M_R - 8$ additions and $136M_R + 112$ multiplications for the preparation stage [1], it is obvious that the overall complexity of the proposed model enjoys a clear advantage.

IV. SIMULATION RESULTS

In order to verify performance of the proposed decoders experimentally, authors simulated a QOSTBC system under four different QAM constellations using four transmit antennas and one receive antenna. Results are tabulated in Tables I and II from which one can deduce complexities of the methods I and II, described in subsections III.A and III.B, as well as those of several state-of-the-art works, by looking at the required number of different operations. For example, when bit error rate (BER) is about $10^{-4}$, complexity of both methods with given QAMs reduce to about 118 addition and 96 multiplication operations, i.e. figures that are comparatively lower than those of [1]-[7] with 16-QAM constellation, hence proving the superiority of the new algorithms. Moreover, Fig. 1 charts BERs of the proposed and ML decoders which illustrates that it is possible to achieve a solution with an optimum performance at lower complexity.

V. CONCLUSION

In this letter, two noble fast maximum-likelihood (ML) decoders for 4×4 QOSTBC were presented. The first method exploited structure of QAM constellations to develop an algorithm that offered less computational complexity than the second approach. Both models initially split the received signal into two parts and detect each one independently. The ML metric of the split signal is then decomposed into a sum of independent positive parts and an interference part. Given that the first parts are independent, the proposed methods substantially limit search area and hence complexity of detection drops exponentially. It was demonstrated that for bit error rate of $10^{-4}$, complexity of the proposed method remains almost unaltered for different QAMs but falls below those of state-of-the-art schemes with 16-QAM.